\documentclass[epsfig]{aa}
\usepackage{epsfig,deluxe}
\begin{document}
 
\newcommand{\gsim}{\hbox{\rlap{$^>$}$_\sim$}}
  \thesaurus{06;  19.63.1}
% A&A Section 6: Form. struct. and  evolut. of stars}
%  \thesaurus{06     % A&A Section 6: Form. struct. and evolut. of stars
%             (03.11.1;  % Cosmogony,
%              16.06.1;  % Planets and satellites: general,
%              19.37.1;  % Stars: formation of,
%              19.53.1;  % Stars: oscillations of,
%              19.63.1)} % Stars: structure of.
%
\authorrunning{A. Dar \& A. De R\'ujula}
\titlerunning{A CB model of GRBs: properties of the $\gamma$-rays}
\title{A Cannonball model of $\gamma$-ray bursts: spectral and temporal
properties of the $\gamma$-rays} 

\author{Arnon Dar$^{1,2}$ and A. De R\'ujula$^1$}
\institute{1. Theory Division, CERN, CH-1211 Geneva 23, Switzerland\\ 
           2. Physics Department and Space Research Institute, Technion,
              Haifa 32000, Israel } 
\maketitle

\maketitle

\begin{abstract} 

Recent observations suggest that gamma ray bursts (GRBs)
and their afterglows are produced by highly relativistic jets emitted in
supernova explosions. We have proposed that the result of the event is not
just a compact object plus the ejecta: within days, a fraction of the
parent star falls back to produce a thick accretion disk. The subsequent
accretion generates jets and constitutes the GRB ``engine'', as in the
observed ejection of relativistic ``cannonballs'' of plasma by
microquasars and active galactic nuclei. Here we investigate the
production of a GRB as the jetted cannonballs exit the supernova shell
reheated by their collision with it, emitting highly forward-collimated
radiation. Each cannonball corresponds to an individual pulse in a GRB. We
cannot predict the timing sequence of these pulses, but the Cannonball
Model fares very well in describing the total energy, energy spectrum, and
time-dependence of the individual pulses.

\end{abstract} 


\section{Introduction}

Once upon a time, Gamma Ray Bursts (GRBs) constituted a sheer mystery,
whose unassailability was reflected in the scores of extremely different
ideas proposed to explain them. In spite of giant strides in the recent
observations ---the discovery of GRB afterglows (Costa et al. 1997; van
Paradijs et al. 1997),  the discovery of the association of GRBs with
supernovae (Galama et al. 1998), and the measurements of the reedshifts
of their host galaxies (Metzger et al. 1997)--- the origin of GRBs is still an
unresolved enigma. In the recent past, the generally accepted view has
been that GRBs are generated by synchrotron emission from fireballs, or
firecones, produced by collapses or mergers of compact stars (Paczynski
1986; Goodman et al. 1987; Meszaros and Rees 1992) by failed supernovae or
collapsars (Woosley 1993; Woosley and MacFadyen 1999; MacFadyen and
Woosley 1999, Woosley 1999) or by hypernova explosions (Paczynski 1998).
But various observations suggest that most GRBs are produced in supernova
events by highly collimated ultrarelativistic jets (Shaviv and Dar 1995;
Dar 1998;  Dar and Plaga 1999; Cen 1999; Dar and De R\'ujula 2000a and
references therein). 

In a previous paper (Dar and De R\'ujula 2000a) we introduced a
relativistic-cannonball 
model in which GRBs are produced by ``cannonballs'' (CBs)
of baryonic plasma emitted subsequently to a core-collapse supernova (SN)
explosion, and are observable when they happen to point
close to our direction. There, we concentrated on GRB afterglows
---due to bremsstrahlung and
synchrotron emission from the CBs after they become transparent to their
own enclosed
radiation--- to emphasize how, in the case of GRB 980425, 
it might be possible to observe the CBs' ``superluminal'' motion.
In this paper we briefly review the CB model and we derive its predictions
for the properties of the $\gamma$-rays in a GRB, generated as the
forward-collimated and blue-shifted
thermal radiation from a succession of 
fast-moving, cooling and expanding CBs ---previously heated
by their collision with the SN shell---
escapes from the transparent outer regions of the shell.
We study the $\gamma$-ray distributions
in time, their energy-spectrum and the correlations between these two
observables, showing that the CB model explains the main observed features
of GRBs.

\section{Jets in astrophysics}

Relativistic jets seem to be emitted by all astrophysical systems wherein
mass is accreted at a high rate from a disk onto a central compact object 
(for a review, see Mirabel and Rodriguez 1999a). 
Highly relativistic jets have been observed in galactic
sources, such as the microquasars GRS 1915+105 (Mirabel and
Rodriguez 1994, 1999a,b;  Rodriguez and Mirabel 1999)
and GRO J165-40 (Tingay et al. 1995) where mass is accreted onto a stellar
black hole, and in many active galactic nuclei  hosting a massive
black hole. These jets are not continuous streams: they consist of
individual blobs of plasma (plasmoids or cannonballs), and their firing 
coincides with a sudden removal of the
accretion-disk material (Belloni 1997; Mirabel and Rodriguez 1999b). 
Cannonballs in microquasars --and presumably also in quasars--
are emitted in pairs, moving in opposite directions.

As they travel, microquasar CBs are observed to expand at a 
speed comparable to, or smaller than, the sound speed of a relativistic
plasma (c/$\sqrt{3}$ in their rest system) probably because the
energy density of their enclosed radiation is comparable to that of their
matter constituency. As they become transparent
and cool down, the CBs' lateral size stabilizes to a roughly constant
value, presumably constrained by magnetic
self-containment and/or by the ram pressure of the ambient material. Quasar
CBs show no measurable expansion as they travel, sometimes for as long as a
million light years (see, e.g., Bridle 2000; Wilson et al. 2000).
Galactic and quasar CBs expand explosively when finally stopped by the
material they traverse.

\section{The cannonball model of GRBs}

The ejection of matter in a supernova (SN) explosion is not fully understood. The
known mechanisms for imparting the required kinetic energy to the ejecta
are inefficient: the theoretical understanding of core-collapse SN events is 
still unsatisfying. It has been proposed (De R\'ujula 1987; Woosley 1993, 
Dar and De R\'ujula 2000a and references therein) 
that the result of a SN event is not just a compact object plus
the ejecta: a fraction of the parent star may be ejected, but
another fraction of its mass may fall back onto the newly born
compact object. For vanishing angular momentum,
the free-fall time of a test-particle from a parent
stellar radius ${\rm R_\star}$ onto an object of mass ${\rm M_c}$ is: 
\begin{eqnarray}
  {\rm t_{fall}}&&{\rm ={\pi\,\left[{R_\star^3\over 8\,G\,M_c}\right]^{1/2} }}
\nonumber \\
&&{\rm \sim 1\; day\; \left[{R_\star\over 10^{12}\;cm}\right]^{3/2}\;
    \left[{1.4\;M_\odot\over M_c}\right]^{1/2}}\, .
\label{tfall}
\end{eqnarray}
The free-fall time is shorter if the mass of the falling material
is not small relative to that of the compact object. 
The fall-time is longer (except for material falling from the polar directions)
if the specific angular momentum is considerably
large, as it is in most stars. The estimate of Eq.(\ref{tfall})
is therefore a rough one.

It is quite natural to suppose that infalling material with non-vanishing
angular momentum settles
into an orbiting disk, or a thick torus if its mass is comparable
to ${\rm M_c}$. We assume that, as observed
in other cases of significant accretion
onto a compact object (microquasars and active galactic nuclei)
in which the infalling material is processed in a series of
``catastrophic'' accretions,  jets of 
relativistic CBs of plasma are ejected. We presume their
composition to be ``baryonic'', as it is in the jets of
SS 433, from which Ly$_\alpha$ and Fe K$_\alpha$ lines have been detected
(Margon 1984), although the violence of the relativistic jetting-process
should in our case break most nuclei into their constituents.

The mechanism producing relativistic jets in accretion
processes and its timing-sequence are not understood
(for suggested possibilities see, e.g., Blandford and Znajek 1977; 
Meszaros and Rees 1997). 
In our model we assume that a series
of CBs is ejected, each one giving rise to one of the
``pulses'' of a specific GRB. After a few pulses the engine
runs out of fuel, and the $\gamma$-ray activity ceases.
The timing sequence of the successive pulses
we are unable to predict, but, as we shall show,
the CB model is quite successful in describing the time-dependence
of the $\gamma$-ray flux {\it within single GRB pulses}.

In brief, the CB model is the following. A sequence of oppositely-directed pairs
of cannonballs is emitted at a time $\rm t_{fall}$ of ${\cal{O}}(1)$ day after
a SN core-collapse. By this time the SN outer shell, traveling at
a velocity $\rm v_S \sim c/10$ (see, e.g., Nakamura et al. 2000)
has moved to a distance:
\begin{equation}
\rm R_S=2.6 \times 10^{14} \;cm\;\left({t_{fall}\over 1\;d}\right)\;
\left({10\,v_S\over c}\right) .
\label{Rs}
\end{equation}
We adopt $\rm R_S=2.6 \times 10^{14}$ cm as a ``reference'' value,
to which our results will be scaled. The reference values of various 
relevant parameters ---that serve as bench-marks to which to scale our results
and imply no strong commitment to their particular choices---
are listed in Table I, for quick reference. We denote with a
bar the actual value of a parameter in the units of its reference value
so that $\rm \overline{R}_S$, for instance, means a given SN-shell radius
divided by $2.6 \times 10^{14}$ cm.

Only if traveling at a small
angle $\theta$ relative to the line of sight, will a CB be
visible. As it hits the SN shell, the CB slows down and heats up.
Its radiation is obscured by the shell
up to a distance of order one radiation length from the shell's
outer surface. As this point is reached, the GRB
is emitted by a CB that
continues to travel, expand and cool down, its radiation being
boosted and collimated by the CB's ultrarelativistic motion.
We do not discuss in this paper the GRB
afterglows (Dar and De R\'ujula 2000a), the flash of  X-ray lines
and the achromatic flare in the afterglow as the electrons and
protons in the GRB recombine  (Dar and De R\'ujula 2000b), 
nor the flux of high energy neutrinos and $\gamma$-rays
produced by the decays of pions made in the CB's collision
with the SN shell (Dar and De R\'ujula 2000c).

There are other events in which a variety of GRBs
could be produced by mechanisms similar to the ones we have
discussed: large mass accretion episodes in binaries including
a compact object, mergers of neutron stars with neutron stars
or black holes (Paczynski 1986, Goodman et al. 1987),
transitions of neutron stars
to hyperon- or quark-stars (Dar 1999; Dar and De R\'ujula, 2000d), etc.
In each case, the ejected cannonballs would make GRBs by
hitting stellar winds or envelopes, circumstellar mass or light.
We discuss only core-collapse SN explosions, as the GRBs 
they would produce
by our mechanism, although relatively ``standard'', satisfactorily
reproduce the general properties of the heterogeneous
ensemble of  GRBs. 

\section{Four ``clocks'' and three energy scales}

Let $\rm \gamma=1/\sqrt{1-\beta^2}={E_{CB}/(M_{CB}c^2)}$ be 
the Lorentz factor
of a CB, that diminishes with time as the CB hits the SN shell
and as it subsequently plows through the interstellar medium. Four
clocks ticking at different paces are relevant to a CB's history.
Let $\rm t_{SN}$ be the
local time in the SN rest system, $\rm t_{CB}$ the time in the CB's 
rest system, $\rm t_{Ob}$ the time   measured by
a nearby observer viewing the CB at an angle $\theta$
away from its direction of motion and $\rm t$ the time
measured by an earthly observer viewing the CB at
the same angle, but from a ``cosmological'' distance 
(redshift $\rm z\neq 0$).
Let x be the distance traveled by the CB in the SN rest system.
The relations between the above quantities are:
\begin{eqnarray}
&&\rm
dt_{SN}=\gamma\,dt_{CB}=\rm{dx\over\beta\, c}\, ;\;\; 
\nonumber \\
&&\rm
dt_{CB}\equiv \delta\,dt_{Ob}\, ;\;\; dt=(1+z)\,dt_{Ob}\;,
\label{times}
\end{eqnarray}
where the Doppler factor $\delta$ is:
\begin{equation}
\rm
\delta\equiv\rm{1\over\gamma\,(1-\beta\cos\theta)}
\simeq\rm {2\,\gamma\over (1+\theta^2\gamma^2)}\; , 
\label{doppler} 
\end{equation}
and its approximate expression is valid for $\theta\ll 1$ and $\gamma\gg 1$,
the domain of interest here.
Notice that for large $\gamma$ and $\theta\gamma$
not large, there is an enormous ``relativistic aberration'':
$\rm dt\sim dt_{SN}/\gamma^2$ and the observer sees
a long CB story as a film in extremely fast motion.
 
The energy of the photons radiated by a CB
in its rest system, $\rm E^\gamma_{CB}$, their energy
in the direction $\theta$
in the local SN system, $\rm E^\gamma_{SN}$,  and the photon
energy, E, measured by a cosmologically distant observer,
are related by:
\begin{equation}
\rm E^\gamma_{CB}=   {E^\gamma_{SN}\over \delta}
\, ;\;\;E^\gamma_{SN}=(1+z)\,E\; ,
\label{energies}
\end{equation}
with $\delta$ as in Eq.(\ref{doppler}).

\section{The making of a GRB}
\subsection{Jet energy and CB mass}

Let ``jet'' stand for the ensemble of CBs emitted in one direction in a SN
event. If a momentum imbalance between the opposite-direction jets is
responsible for the large peculiar velocities ${\rm v_{NS}\approx 450\pm
90~ km~s^{-1}}$ (Lyne and Lorimer 1994) of neutron stars born in SNe, the
jet kinetic energy $\rm E_{jet}$ must be, as we shall assume for our GRB
engine, larger than $\sim 10^{52}$ erg (e.g. Dar and Plaga 1999). The
jet-emitting process may be ``up-down'' symmetric to a very good
approximation, in which case the jet energies may be much bigger. There is
evidence that in the accretion of matter by black holes in quasars
(Celotti at al. 1997; Ghisellini 2000) and microquasars (Mirabel and 
Rodriguez 1999a,b) the
efficiency for the conversion of gravitational binding energy into jet
energy is surprisingly large. 
If in the production of CBs the central compact object
in a SN ingurgitates several solar masses, it is not
out of the question that $\rm E_{jet}$ be as large as
$\rm M_\odot c^2\sim 1.8\times 10^{54}$ erg. We shall
adopt here a compromise value, $10^{53}$ ergs, as the reference
jet energy.

Let $\rm \gamma_{in}$ be the Lorentz factor of a cannonball
as it is fired. Let $\rm E_{CB}=f\,E_{jet}$ be the energy of a CB;
on average GRBs have some five to ten significant pulses, so that the fraction 
f may typically be 1/5 or 1/10. We shall adopt $\rm E_{CB}=10^{52}$ erg
as our reference value. For this value, the CB's mass is
comparable to an Earth mass:
%${\rm M_{jet}\sim 1.5 \times 10^{-6}\,M_{NS}\,(10^3/\gamma)}$,
${\rm M_{CB}\sim 1.8\, M_\otimes (10^3/\gamma_{in})}$, for a Lorentz factor
of $\rm\gamma_{in}={\cal{O}}(10^3)$, that we shall find to be ``typical''.

\subsection{CB deceleration by the SN shell}

Let $\rm \beta_{in}\, c$ be the expansion velocity of a CB,
in its rest system, as it travels from the point of emission to
the point at which it reaches the SN shell. For the reference
value of $\rm \beta_{in}$, as reported in Table I, we use $1/(10\,\sqrt{3})$: 
one tenth of the sound velocity of a relativistic plasma.
The CB reaches the shell with a radius   
\begin{equation}
\rm R_{CB}\sim R_S\,{\beta_{in}\over \gamma_{in}}
\label{radiusCB}
\end{equation}
 and sweeps
up a ``target'' mass 
$\rm M_T\sim\pi\,R_{CB}^2\,X_S=M_S\,\beta_{in}^2/(4\,\gamma_{in}^2)$,
or some $\rm \sim 2.8\times 10^{-3}\,M_\otimes$, for our reference
parameter values and  $\rm\gamma_{in}=10^3$.
The CB and the SN shell are ``thick'' in the sense of extending over
many radiation lengths and many nucleon-nucleon interaction lengths.
A high-energy  nucleon suffering successive interactions in a dilute
gas or plasma loses roughly 2/3 of its energy to $\pi^\pm$
production, with most of the pion energy being
carried away by the neutrinos in $\pi\to \mu\,\nu$ decays
and the subsequent $\mu$ decays. The electrons
from $\mu$  decay and the photons from $\pi^0$ decay locally
deposit roughly 1/3 of the original nucleon energy. 

The Lorentz factor of the CB after it has swept the SN shell,
$\rm \gamma_{out}$, is simply the 
ratio of the total energy to the invariant mass 
($\rm \sqrt{s}=M\,c^2$) of the outgoing object:
\begin{equation} 
\rm \gamma_{out}\simeq {E_{CB}/3\over \sqrt{s}}
\simeq{E_{CB}/3\over
\sqrt{2\,M_T\,c^2\,E_{CB}/3+M^2_{CB}\,c^4}}\; ,
\label{gammaout1}
\end{equation}
where we have  used  $\rm E_{CB}\gg M_Tc^2$. Substituting for
$\rm M_T$ and $\rm M_{CB}$ as functions of $\rm\gamma_{in}$
and $\rm\beta_{in}$, one obtains:
\begin{equation} 
\rm \gamma_{out} \simeq \gamma_{in} 
\;\sqrt{2\,E_{CB}\over 3\,\beta_{in}^2\,M_S\, c^2+18\,E_{CB}}
\label{gammaout}
\end{equation}
whose limiting values are:
\begin{eqnarray}
&& \rm \gamma_{out}\sim {\gamma_{in}\over 3}\;\;\;\; 
(for\; 6\,E_{CB}\gg \beta_{in}^2\,M_S\,c^2)\, , \nonumber\\ 
&& \rm \gamma_{out}
\sim {\gamma_{in}\over 3\,\overline{\beta}_{in}}\,
\left[{\overline{E}_{CB}\over \overline{M}_S}\right]^{1\over 2}\;\;\;\;
(for\; 6\,E_{CB}\ll \beta_{in}^2\,M_S\,c^2)\; .
\label{gammaout2}
\end{eqnarray}
For our reference $\rm \gamma_{out}\sim 10^3$,
the  values of $\rm \gamma_{in}$ implied by Eqs.(\ref{gammaout2})
may look surprisingly large. But Eqs.(\ref{gammaout2}) do not
depend on $\rm R_S$: any relativistic jet exiting from the core of a SN
encounters the same amount of non-collapsed material, and must have a 
$\rm \gamma_{in}$ considerably larger than $\rm\gamma_{out}$.

The very large value of $\rm\gamma_{in}$ ($\sim 3 \times 10^3$ for our reference
parameters) implies that the fractional solid angle covered by a CB as it hits the
SN shell is tiny: $\rm \beta_{in}^2/(4\,\gamma_{in}^2)\sim 10^{-10}$,
again for our reference parameters.  This presumably makes it unlikely for 
successive CBs to hit precisely the same spot in the SN shell:
CB-CB collisions and mergers may be the exception, rather than the rule.

\subsection{Attenuation of the $\gamma$ rays}

The density profile of the outer layers of a SN shell as a function
of the distance x to the SN center can be measured from the photometry,
spectroscopy and evolution of the SN emissions (see e.g. Nakamura et al. 2000 
and references 
therein). The observations can be fit by a power law,
$\rm x^{-n}$, with $\rm n \sim 4\; to\, 8$. Our results are sensitive to
this density profile only in the outer region where the SN shell
becomes transparent (and the measurements are made), so that we
can adopt the same profile at all $\rm x>R_S$:
\begin{equation}
\rm \rho(x)=\rm\rho(R_S)\,\Theta(x-R_S)\,\left[{R_S\over x}\right]^n\, .
\label{profile}
\end{equation}
The SN-shell grammage still in front of a CB located at x is:
\begin{equation}
\rm X_S(x)=\int_x^\infty \, \rho(y)\,dy=
{M_S\over 4\,\pi\, R_S^2}\; \left[{R_S\over x}\right]^{n-1}\, .
\label{SNgram}
\end{equation}

For photons in the MeV domain the attenuation length is similar, within
a factor 2, in all elements from H to Fe (Groom et al., 2000), and can be 
roughly approximated by:
\begin{equation} 
\rm X_\gamma(E)\sim 1.0\,(E/keV)^{0.33}\; g\, cm^{-2}\; .
\label{Xgamma}
\end{equation}
The value of $\rm X_\gamma(E)$ in the $\rm E=10$ keV to 1 MeV domain
(2.1 to 9.8 gr/cm$^2$) is close to the attenuation length in a hydrogenic
plasma ($\rm X_\gamma^{ion}\simeq m_p/\sigma_{_T}\simeq 2.6$ gr/cm$^2$,
with $\rm m_p$ the proton's mass and
$\rm \sigma_{_T}\simeq 0.65\times 10^{-24}$ cm$^2$ the Thomson 
cross-section). Therefore, it makes little difference in practice whether
or not we take into account that the SN-shell material reached
by the CB may be ionized by its previously emitted radiation.
Equating $\rm X_S(x)=X_\gamma(E)$ and solving for x, we define a useful
quantity:  $\rm x_{tp}(E)$, the position at which the SN shell becomes
(one-radiation-length) transparent:
\begin{equation}
\rm x_{tp}(E) = R_S\;\left[{M_S\over 4\,\pi\, R_S^2}\;
{1\over X_\gamma(E)}\right]^{1\over n-1}\propto E^{-0.33/(n-1)}\; ,
\label{SNtransparent}
\end{equation}
whose energy dependence is extremely weak.
Blue-shifted to the SN rest-system, as in Eq.(\ref{energies}),
GRB photons have energies
in the MeV range. Let $\rm \tilde x_{tp}\equiv x_{tp}(1\; MeV)$.
 For  our reference parameters, some representative results are: 
$\rm \tilde x_{tp}\simeq 2.9\,R_S$ for $\rm n=8$,
$\rm \tilde x_{tp}\simeq 4.5\,R_S$ for $\rm n=6$. 
At $\rm \tilde x_{tp}$, $\rm \rho(x)$
is orders of magnitude smaller than at $\rm x\sim R_S$, where most of
the SN-shell's mass is steeply concentrated. This will simplify our
discussion, for it is a fair approximation to have the CB
slow down at heat up close to $\rm x=R_S$, and proceed thereafter unperturbed
by the SN-shell material, except inasmuch as little of its radiation
can escape before it reaches $\rm x=\tilde x_{tp}$. At that point, the CB
has expanded from the radius $\rm R_{CB}$ of Eq.(\ref{radiusCB}) to
a radius at transparency:
\begin{equation}
\rm R_{CB}^{tp}\simeq R_{CB}+
{\tilde x_{tp}-R_S\over \gamma_{out}}\;\beta_{out}
\simeq 
{\tilde x_{tp}-R_S\over \gamma_{out}}\;\beta_{out}\; , 
\label{Rtrans}
\end{equation}
some $2.9\times 10^{11}$ cm, for our reference parameters.
The CB itself becomes transparent to the radiation it encloses later,
when it reaches a radius
$\rm \widetilde R^{tp}_{CB}\simeq [3\,
M_{CB}\,\sigma_T/(4\,\pi\,m_p)]^{1\over 2}$.
We expect the CB to stop expanding at a proper
quasi-relativistic rate $\rm\beta_{out}$  soon after it becomes
transparent and its inner radiation pressure drops abruptly: the inertial
mildly relativistic
transverse motion of its matter constituents is slowed-down by
interstellar material and, perhaps, by self-confining magnetic fields.

\subsection{Total energy of a GRB pulse}

A CB expanding as a quasi-relativistic plasma ought to reach
the SN shell with a shape (in its rest system) very close to spherical.
The microscopic description of what happens as the CB and the
material of the SN shell collide and coalesce is elaborate
(Dar and De R\'ujula, 2000c).
Much of the available energy is deposited at the CB's front surface
by nucleons sharing their energy and $\gamma$'s from
$\pi^0$-decay depositing theirs. Electrons from $\mu$ decay deposit 
their energy much deeper into the CB. The ionized CB's material
is hot and dense enough for the deposited energy to thermalize very fast. 
As it impinges the SN shell, a CB may have a tendency to get flattened,
but the velocities, $\beta\approx 1-1/(2\gamma^2)$,
corresponding to $\rm\gamma_{in}$ and $\rm\gamma_{out}$
are so similar that no significant flattening occurs between
the time the CB hits the shell and the time it reaches
the point at which the shell becomes
transparent to the CB's radiation: flattening is subdominant relative
to the CB's expansion. For the subsequent estimates we approximate
the CB as  a spherical body with a uniform internal temperature.

The proper temperature $\rm T_0$ acquired by the CB as it hits the SN shell
is high enough for the CB's internal-radiation 
energy-density to be much larger 
than the mass-energy density of its matter constituents. Consequently,
$\rm T_0$ can be estimated by equating the total internal radiation
energy to the invariant mass in the CB-SN shell collision:
\begin{equation}
\rm T_0\simeq
\left[{3\over 8\,\pi\, a}\,\sqrt{3\over 2}\;
{(3\,\beta_{in}^2\,M_S\,c^2+18\,E_{CB})^{3/ 2}
\over \sqrt{E_{CB}}}\;{\gamma_{out}^2\over R_S^3}\right]^{1\over 4}
 ,
\label{T0}
\end{equation}
where $\rm a\simeq 1.37 \times 10^{14}$ erg cm$^{-3}$ keV$^{-4}$
is the radiation-density constant.
The result, $\rm T_0\sim 3.4$ keV for our reference parameters,
is very insensitive to their values,
but for the $\rm R_S^{-3/4}$ dependence on the SN-shell's radius.

Since the SN-shell's material is highly concentrated close to $\rm x=R_S$,
as in Eq.(\ref{SNgram}), we can take $\rm T_0$ to be the temperature
at that point. A rough estimate of the total energy in a GRB pulse can
be obtained as follows. 
While on the part of the shell that is not transparent,
the CB does not lose much energy via surface radiation, so that it 
expands quasi-adiabatically
at roughly constant $\rm R_{CB}(t)\,T(t)$. At the point
at which the shell becomes transparent, the internal-radiation energy in the 
CB is reduced, from the value $\rm \sqrt{s}$ of Eq.(\ref{gammaout1}), to 
$\rm E_{tp}\simeq \sqrt{s}\; R_{CB}/R_{CB}^{tp}$.
Approximately 1/e of this energy is emitted thereafter, its value in the
CB's rest system is:
\begin{equation}
\rm E_{pulse}^{rest}\simeq {1\over 3\,e}\;{E_{CB}\over \gamma_{in}}\;
{R_{S}\over \tilde x_{tp}-R_S}\;
{\beta_{in}\,\gamma_{out}\over \beta_{out}\,\gamma_{in}}\; ,
\label{EGRB}
\end{equation}
whose limiting values are:
\begin{eqnarray}
&&\rm E_{pulse}^{rest}\simeq 
4.5\times 10^{45}\; erg\,\left[{3\,R_S\over \tilde x_{tp}-R_S}\right]\,
{1\over \overline{\gamma}_{out}}\,
{\overline{\beta}_{in}\over\overline{\beta}_{out}}\nonumber \\
&&\rm \;\;\;\;\;\;\;\;\;\;\;
(for\; 6\,E_{CB}\gg \beta_{in}^2\,M_S\,c^2)\, ,
\nonumber\\
&&\rm E_{pulse}^{rest}\simeq 
4.5\times 10^{45}\; erg\,\left[{3\,R_S\over \tilde x_{tp}-R_S}\right]\,
{1\over \overline{\gamma}_{out}\,\overline{\beta}_{in}\,\overline{\beta}_{out}}\,
{\overline{E}_{CB}^2\over \overline{M}_S}
\nonumber\\
&&\rm \;\;\;\;\;\;\;\;\;\;\;
(for\; 6\,E_{CB}\ll \beta_{in}^2\,M_S\,c^2)\;  .
\label{EGRB1}
\end{eqnarray}

 An observer at rest,
located at a known luminosity distance $\rm D_L(z)$ from the CB and
viewing it at an angle $\theta$ from its direction of motion would measure
a  ``total'' (time- and energy-integrated) fluence per unit area:
\begin{equation}
\rm {dF\over d\Omega}\simeq {1+z\over 4\,\pi\,D_L^2}
\,{E_{pulse}^{rest}}\;\delta^3\; ,
\label{dfdomega}
\end{equation}
where $\delta=\delta[\gamma,\theta]$ is given, here and in what follows, 
by Eq.(\ref{doppler})
with $\rm \gamma=\gamma_{out}$.
The ``spherical'' energy deduced from this result would be an overestimate
of the true energy $\rm E_{pulse}^{rest}$
by the last factor in Eq.(\ref{dfdomega}), which,  for $\rm \gamma_{out}=10^3$ 
and $\theta\gamma\sim {\cal{O}}(1)$, can be as large as $\sim 10^9$.
Enhanced by a factor ranging up to this large number, the GRB-pulse energies of 
Eq.(\ref{EGRB}) can easily reproduce the observations, as discussed
in detail in Section 7.

Armed with an expression such as Eq.(\ref{dfdomega}) one can embark in the
exercise of studying the extent to which GRBs are standard candles, by 
checking whether the observations at fixed redshift are statistically compatible 
with this expression for a uniformly distributed $\cos\theta$ distribution.
But the current number of GRBs with measured redshifts is only fifteen,
and their deduced total energies are affected by absorption, by experimental
efficiency and threshold effects, etc. We do not, in this paper, attempt such
an analysis, that has been initiated, with encouraging results, by Plaga (2000),
who uses ---to extract redshifts from a large collection of GRBs--- the
``Cepheid-like'' relationship between variability and luminosity proposed
by Fenimore and Ramirez-Ruiz (2000).

\subsection{Energy and time dependence of a $\gamma$-ray pulse}

A CB, as it reaches the transparent outskirts of a SN shell, is
expanding and cooling and its radiation is becoming visible to
the observer.  In what follows it is convenient to measure the 
GRB observer's time,
t, setting $\rm t=0$ at the moment of the encounter of the CB and the
SN shell. The time of (one-radiation-length) transparency is then:
\begin{equation}
\rm t_{tp}\simeq  {1+z\over \gamma_{out}\,\delta}\;
{\tilde x_{tp} - R_S\over c}\; .
\label{tttp}
\end{equation}
The CB temperature at $\rm t= t_{tp}$ is:
\begin{equation}
\rm T_{tp}\sim\rm\left[{3\over 4\,\pi\,a}\;
{E_{pulse}^{rest}\over (R_{CB}^{tp})^3}
\right]^{1\over 4}\; ,
\label{Ttrans}
\end{equation}
whose limiting values are:
\begin{eqnarray}
&&\rm T_{tp}\sim 0.1\;keV \left[{3\,R_S\over \tilde x_{tp}-R_S}\right]\,
{\overline{\gamma}_{out}^{1\over 2}\,
\overline{\beta}_{in}^{1\over 4}\over\overline{\beta}_{out}}\nonumber\\
&&\rm \;\;\;\;\;\;\;\;\;\;\;
(for\; 6\,E_{CB}\gg \beta_{in}^2\,M_S\,c^2)\, ,
\nonumber\\
&&\rm T_{tp}\sim 0.1\;keV \left[{3\,R_S\over \tilde x_{tp}-R_S}\right]\,
{\overline{\gamma}_{out}^{1\over 2}
 \over \overline{\beta}_{in}^{1\over 4}\,\overline{\beta}_{out}}\;
{\overline{E}_{CB}^{1\over 2}\over \overline{M}_S^{1\over 4}}
\nonumber\\
&&\rm \;\;\;\;\;\;\;\;\;\;\;
(for\; 6\,E_{CB}\ll \beta_{in}^2\,M_S\,c^2)\; .
\label{Ttrans1}
\end{eqnarray}

The time-dependences
of the CB's radius, its temperature, and the distance x of the CB from the
SN's center are, for $\rm t>0$:
\begin{eqnarray}
&&\rm R_{CB}[t]\simeq R_{CB} + R_{CB}^{tp}\,{t\over t_{tp}}
\sim R_{CB}^{tp}\,{t\over t_{tp}}
\, ,\nonumber\\
&&\rm T[t]\simeq \rm T_{tp}\;{R_{CB}^{tp}\over R_{CB}[t]}\, ,\nonumber\\
&&\rm x[t]\simeq R_S+{\delta \,\gamma_{out}\over 1+z}\;c\, t\;.
\label{timedeps}
\end{eqnarray}

Let the number of photons per unit time and energy, assumed to be
isotropically emitted by the CB in its rest system, be:
\begin{equation}
\rm {dn_\gamma\over dE_\gamma\;dt_{CB}}\equiv F(E_{CB}^\gamma,T)\, .
\label{restemission}
\end{equation}
Using Eqs.(\ref{times}-\ref{energies}) to change variables to
$\rm E$ and t (the $\gamma$-ray energy and time in the
observer's frame), we obtain:
\begin{equation}
\rm {dn_\gamma\over dE\,dt}\simeq
F\left(E\,{1+z\over\delta},T[t]\right)\; .
\label{boostany}
\end{equation}
In the approximation in which the CB's emission in its rest system
is a thermal distribution from its surface, the function F is:
\begin{equation}
\rm  F(E_{CB}^\gamma,T) \simeq
 {2\,\pi\,\sigma\over \zeta(3)}\;(R_{CB}[t])^2\;
{(E_{CB}^\gamma)^2\over Exp\{{E_{CB}^\gamma/T}\}-1}\; ,
\label{thermal}
\end{equation}
where $\rm \sigma=c\,a/4$ is the Stefan-Boltzmann constant.

The observed energy and time dependence of the photon intensity 
(photon number per unit area, N) 
of a single pulse  in a GRB at an angle $\theta$ relative to the CB's
motion is then predicted to be:
\begin{eqnarray}
&&\rm {dN\over dE\,dt}\equiv {1+z\over 4\,\pi\,D_L^2}\;
\delta^2\,  {dn_\gamma\over dE\,dt}\, ,\\
&&
\rm {dn_\gamma\over dE\,dt}\simeq {2\,\pi\,\sigma\over \zeta(3)}\;
{\left[R_{CB}[t]\;E\,(1+z)/\delta\right]^2\; Abs(E,t)\over 
Exp\left\{E\,(1+z)/(\delta\, T[t])\right\}-1}\; ,
\label{boostthermal}
\end{eqnarray}
with $\rm R_{CB}[t]$ and $\rm T[t]$ as in Eqs.(\ref{timedeps}), 
and where
\begin{equation}
\rm Abs(E,t)=
Exp\left[-{X_S(x[t])\over X_\gamma(E\,(1+z))}\right]
\label{attenuation}
\end{equation}
is the attenuation of the flux in the SN shell.

For n in Eq.(\ref{SNtransparent}) as large as the observations
indicate ($\rm n\sim 6$), the absorption factor $\rm Abs(E^\gamma,t)$
rises very sharply from 0 to 1 around $\rm t=t_{tp}$,  in which case
the width of a GRB pulse in energy and time is governed by
the exponential in the denominator of
Eq.(\ref{boostthermal}). The argument of that exponential
can be simply rewritten as $\rm E\,t/H$, with:
\begin{equation}
\rm H\equiv\rm {\tilde x_{tp}-R_S\over c\,\gamma_{out}}\, T_{tp}\; ,
\label{h}
\end{equation}
whose limiting values are:
\begin{eqnarray}
&&\rm H\sim 2.5\; keV\, s\; 
{ \overline{\beta}_{in}^{1\over 4}\over 
\overline{\gamma}_{out}^{1\over 2}\,\overline{\beta}_{out}}
\nonumber\\
&&\rm \;\;\;\;\;\;\;\;\;\;\;
(for\; 6\,E_{CB}\gg \beta_{in}^2\,M_S\,c^2)\, , 
\nonumber\\
&&\rm H\sim 2.5\; keV\, s\; 
{ 1 \over 
\overline{\gamma}_{out}^{1\over 2}\,\overline{\beta}_{out}\,
\overline{\beta}_{in}^{1\over 4}}\;
{\overline{E}_{CB}^{1\over 2}\over\overline{M}_S^{1\over 4}}
\nonumber\\
&&\rm \;\;\;\;\;\;\;\;\;\;\;
(for\; 6\,E_{CB}\ll \beta_{in}^2\,M_S\,c^2)\; .
\label{h1}
\end{eqnarray}

\section{Some simplifications and approximate correlations}

To guide the  eye, we give a simplified
approximate form of Eq.(\ref{SNtransparent}), which we do not use
in our explicit calculations:
\begin{equation}
\rm {dN\over dE\,dt}\propto
{(E\,t)^2\over Exp\{E\,t/H\}-1}\,
Exp \left\{-\left[ {t_{tp}/ t}\right]^{n-1}\right\}
\;\Theta[t]\; .
\label{simple}
\end{equation}
The total photon intensity and energy flux are, in this 
approximation:
\begin{equation}
\rm {dN\over\,dt}\propto
%{dI_\gamma\over dt}(t_{tp}) \,
\Theta[t]\; {t_{tp}\over t}\, Exp 
\left\{-\left[ {t_{tp}/ t}\right]^{n-1}\right\}\,, 
\label{simple2}
\end{equation}
\begin{equation}
\rm {F_E(t)}\propto
%F_\E(t_{tp})\,
\Theta[t]\; \left[{t_{tp}\over t}\right]^2\,
Exp \left\{-\left[ {t_{tp}/ t}\right]^{n-1}\right\}\,.
\label{simple3}
\end{equation}
Let the peak $\gamma$-ray 
energy at a fixed time during a GRB pulse be defined as
$\rm E^\gamma_p(t) \equiv max\,[ E^2\,dI_\gamma/dE\, dt]$.
Its value is   $\rm E^\gamma_p(t)\simeq 3.92\,\delta\,T[t]/(1+z) $, so that,
for t near or after $\rm t_{tp}$:
\begin{equation}
\rm E^\gamma_p(t) \simeq E^\gamma_p(t_{tp})
\;\Theta[t]\; {t_{tp}\over t}\,.
 \label{simple4}
\end{equation}

The total ``isotropic'' energy of a GRB pulse ---deduced from its observed 
fluence assuming an isotropic emission--- can be deduced from
Eq.~(\ref{dfdomega}), to be:
\begin{equation}
\rm   E_{iso}=
{4\,\pi\,D_L^2\, F \over 1+z}\simeq E_{pulse}^{rest}\, \delta^3\,.
\label{eisot}
\end{equation}

If CBs were  ``standard candles'' with fixed mass, energy
and velocity of expansion,
and if all SN shells had the same 
mass, radius and density distribution, all differences between 
GRB pulses would result from their different distances and angles
of observation. For such standard candles 
it follows from Eqs.(\ref{times}-\ref{energies},\ref{eisot}) that the observed
durations (half widths at half maximum)  of the photon intensity and 
of the energy flux density ($\rm \Delta t_I$ and
 $\rm \Delta t_F$),
their peak values  ($\rm N_p$ and $\rm F_p)$, and the peak energy 
($\rm E^\gamma_p$) in a single GRB pulse are 
roughly correlated to the total ``observed'' isotropic energy 
($\rm E_{iso}$) as follows:
\begin{equation}
\rm \Delta t_I\propto (1+z)\, [E_{iso}]^{-1/3}\,,  
\label{twidthi}
\end{equation}  
\begin{equation}
\rm \Delta t_F\propto (1+z)\, [E_{iso}]^{-1/3},  
\label{twidthf}
\end{equation}  
\begin{equation}
\rm N_p\propto E_{iso},  
\label{Ipeak}
\end{equation}  
\begin{equation}
\rm F_p\propto[ E_{iso}]^{4/3}\, (1+z)^{-1}\, , 
\label{Lpeak}
\end{equation}  
\begin{equation}
\rm E^\gamma_p\propto [E_{iso}]^{1/3}\,(1+z)^{-1}\, .  
\label{Epeak}
\end{equation}  
These approximate correlations can be tested using the sample of 15 GRBs with
known redshifts. 
Because of the strong dependence of the CB pulses 
on the Doppler factor and their much weaker dependence on the 
other parameters, they may be approximately satisfied
(see, e.g. Plaga 2000) in spite of the fact 
that CBs and SN shells are likely to be sufficiently varied
not to result in standard candles. 

Within the standard-candle approximation there is also a simple
correlation between the rate and the fluence of GRBs. For the 
region of the universe that is close enough to us to be approximately
homogeneous and Euclidean, Eq.(\ref{dfdomega}) implies that
$ \rm F=E_{rest}^{pulse}\, \delta^3 /(4\, \pi\, D^2)$.
If CBs were {\it stationary}, the
corresponding rate of GRB pulses with fluence larger than 
a given $\rm F_0$ 
would satisfy the well known relation:
\begin{equation}
\rm \dot N(>F_0)\simeq \dot n_{CB} {4\,\pi\over 3}\,
               \left[{E_{rest}^{pulse}\over 4\,\pi\,F_0}\right]^{3\over 2}\propto 
F_0^{-3/2}, 
\label{euclid1}
\end{equation} 
where $\rm \dot n_{CB}$ is the mean production rate of 
 CBs per unit volume. For our
highly relativistic CBs, whose ``isotropic'' energy  
is multiplied by the factor $\rm \delta^3$, Eq.(\ref{euclid1}) 
is modified to:
\begin{equation}
\rm \dot N(>F_0)\simeq {3\over 7}\, {2^{7\over 2}\over\gamma^2}
\,\dot n_{CB}\,
  {4\,\pi\over 3}\,\left[{\gamma^3\, E_{rest}^{pulse}\over 
4\,\pi\,F_0}\right]^{3\over 2}
  \propto  F_0^{-3/2}, 
\label{euclid2}
\end{equation} 
yielding the same rate-to-fluence relation. The same power-law scaling
is obtained for the relation between the rate and the
peak-energy density-flux from CBs. Both relations
should be approximately satisfied by  very bright (relatively nearby) GRBs.

For distant GRBs the above relations are sensitive to the cosmological
model, to the not-well-determined SN- (or star-formation) rate and to
the strong selection effect favouring observations of distant GRBs with
large $\gamma$ and a small viewing angle $\theta$, and a correspondingly 
large Doppler-enhancement $\delta$. Because of this, we do 
not discuss here the rate-to-fluence relation for 
faint GRBs (Yi, 1994; Plaga 2000).
  
\section{Predictions of the Cannonball Model}

Some common properties of GRB pulses 
(for detailed light curves see Kippen 2000; Mallozzi 2000) are observed 
to be: 
\begin{itemize}
\item{(a)} The GRB fluences, integrated in energy and time,
lie within one or two orders of magnitude above or below
10$^{-5}$ erg/cm$^2$ (see, e.g., Paciesas et al. 1999).
\item{(b)} Individual pulses are narrower in time, the higher the
energy interval of their individual photons
(see, e.g., Fenimore et al. 1995).
\item{(c)} Individual pulses rise and peak at earlier time, the higher the
energy interval of their individual photons
(see, e.g., Norris et al. 1999; Wu and Fenimore 2000)
\item{(d)} Individual pulses have smaller photon energies, the
later the time-interval of observation (see, e.g., Preece et al. 1998) .
\item{(e)} The energy spectrum of GRBs, or of their individual
pulses, if plotted as $\rm E^2\,dN/dE$, rises with energy as $\rm E^\alpha$,
with $\alpha \sim 1$,  has a broad peak at $\rm E\sim 0.1$ to 1 MeV, and 
decreases thereafter (see, e.g., Preece 2000).
\item{(f)} Most GRBs consist of pulses whose time-behaviour is 
a fast rise followed
by an approximately exponential decay: a ``FRED'' shape. Some
GRBs have non-FRED, roughly  time-symmetric pulses
(see e.g., Fenimore et al. 1995 and references therein) 
The overwhelming majority of GRBs are either made of FRED or non-FRED
pulses: there are no GRBs with mixed pulse-shapes.
\end{itemize}

All of the above items are properties of the CB model.

In Fig.(\ref{flu}) we illustrate item (a) by plotting the total
fluence, estimated with use of Eq.(\ref{dfdomega}) and
varying one parameter at a time. Naturally, the highest sensitivity
is that to the viewing angle $\theta$, followed by that to
$\rm E_{CB}$ and z.
The remaining itemized GRB properties
all follow from Eq.(\ref{boostthermal}); 
items (b,c,d) are even apparent in the simplified Eq.(\ref{simple})
for the time and energy dependence of the $\gamma$-ray flux.
In Fig.(\ref{t3Es}) we illustrate items (b) and (c) by plotting 
Eq.(\ref{boostthermal}) at three fixed $\gamma$-ray energies, for all parameters 
fixed at the reference values of Table I. In Fig.(\ref{E3ts}) we
similarly illustrate item (d) in a plot at three different times, multiples
of the time of shell transparency. Item (e) is illustrated in Fig.(\ref{E})
where we plot $\rm E^2\, dN/dE$, obtained by integrating
Eq.(\ref{boostthermal}) over all times; the figure also reports the
sensitivity
to various parameters, by modifying them, one at a time, relative
to the reference parameters.  In Fig.(\ref{t}) we
illustrate item (f) by plotting $\rm dN/dt$, obtained by integrating
Eq.(\ref{boostthermal}) over all energies above 30 keV. Once
again, we vary reference parameters as in Fig.(\ref{E}).
Redshift  not being a free parameter specific to our model,
we separately illustrate in Fig.(\ref{z}) the z-dependence of the time-integrated
and energy-integrated versions of Eq.(\ref{boostthermal}).

A look at Figs.(\ref{t}) and (\ref{z}) reveals that, for the parameter 
ranges explored
therein, all the predicted GRB-pulse shapes are FREDs
and are relatively short in time (fractions of a second). Yet, these are not
general predictions of Eq.(\ref{boostthermal}). It is, for instance, quite
conceivable that the ejection of a shell in a SN explosion be due to
one or various CBs emitted immediately after core implosion: the shock
wave induced by their passage through the outer shells of the star
would trigger their ejection. In that case the outgoing shell would be
quite disrupted in the ``polar'' directions in which later CBs would 
result in a GRB. It is also possible that a GRB be due to the passage
of CBs through material expelled by a parent-star's wind, as opposed
to the SN shell. In both cases, the density profile of the matter traversed
by a GRB may be very different from that described by a large index
$\rm n\sim 4$ to 8 in Eq.(\ref{profile}), indicated by observations of
complete SN shells, not of their small polar regions.
In Fig.(\ref{t2}) we illustrate these points by plotting
$\rm dN/dt$ for $\rm n=2,\,3$, with the rest of the parameters
at their reference values, and we also give an example with
$\rm n=6$ and a very large viewing angle $\rm \theta=20/\gamma_{out}$.
All three of these time-profiles are quite symmetrical non-FREDs
and have durations in the few-second range (it is also possible
to generate long-duration FREDs, as we shall see below in the 
specific case of GRB 980425).

In  Fig.(\ref{6CB}) we plot a GRB with 6 CBs, shot at random times
in a 1.5 s interval and with random values of $\rm E_{CB}$ within
a factor of three of our reference value. All other parameters
in this figure, but the SN-shell density-profile index n,
have their reference values: the only difference
between Fig.(\ref{6CB}a) and  Fig.(\ref{6CB}b) is that
$\rm n=8$ in the former, $\rm n=4$ in the latter.
These figures illustrate the obvious
fact that the correspondence between CBs and observed pulses
need not be biunivocal: a CB produces a GRB pulse, but an observed
pulse can be due to a superposition of CB subpulses. Notice that this is also
a way to obtain pulses that are very wide, or do not have FRED- 
or symmetrical shapes.

\section{Brief comparison to some data}

Comparing a GRB theory with specific GRBs is a tricky task, for an
obvious reason: GRBs being all different, one may
be tempted to choose GRBs that fit the theory,
rather than doing the opposite. In this Section we investigate three
GRBs with measured redshifts. Of this ensemble, we use 
the highest fluence event (Briggs et al 1999)
GRB 990123 ($\rm z=1.6$),  to analize the energy 
distribution; we use GRB 980425 (Kippen 2000), 
whose redshift (Galama et al. 1998)  is by far the smallest
($\rm z=0.0085$) and yet has a conventional fluence
(Kippen et al. 1998), to study
the time-dependence of its single pulse;  finally, we use GRB 990712
($\rm z=0.4315$) to study the correlation between the 
$\gamma$-ray energy- and time-distributions (Mallozzi 2000)
and to expose the limitations of the
CB model in its present simple form.

\subsection{Energy dependence}
The predicted energy spectrum of a GRB is obtained by integrating
Eq.(\ref{boostthermal}) over all times. The resulting flux 
distribution and the same result weighed with $\rm E^2$ are compared with 
the GRB 990123 data in Fig.(\ref{123}). The parameters used
are $\rm\overline\beta_{in}=1/4$, $\rm\overline\beta_{out}=1$,
$\rm\overline{M}_S=1/5$, $\rm\overline{R}_S=1/2$, 
$\rm\overline{E}_{CB}=20$, $\rm\overline{\gamma}_{out}=1.5$,
$\rm\overline{\theta}_S=1.46$, $\rm n=6$ and $\rm z=1.6$.
The value of $\rm E_{CB}$ may look large, but this is a 
multiple-pulse GRB and the energy distribution is integrated
over all pulses: $\rm\overline{E}_{CB}=20$ corresponds only
to twice our reference value for $\rm E_{jet}$. For these parameters
the GRB fluence, as estimated via Eq.(\ref{dfdomega}), is the
observed 26.5 10$^{-5}$ erg cm$^{-2}$. Since the shape of
the energy distribution is insensitive to the various parameters,
as seen in Fig.(\ref{E}), it is easy to find many parameter ensembles
that result in the same prediction: the energy distribution by itself
is not a good observable to constrain the input, but is, on the other
hand, a solid test of the model.

The shape of the energy spectrum $\rm dN/dE$ of Fig.({\ref{123}a)
can be easily understood. At the lower energies, the $\sim$1/E behaviour
is the result of integration over thermal spectra with temperatures
that decrease with time, see Fig.(\ref{E3ts}). The abrupt decrease
of $\rm dN/dE$
at the higher energies reflects the input thermal spectrum at
the time and
temperature at which the SN shell starts to become transparent.

The comparison made in Fig.(\ref{123}) is quite satisfactory, particularly
if one realises that many of the higher-energy data are but upper limits.
In making this figure we used the thermal distribution of
Eq.(\ref{thermal}), and the fact that at the higher energies the theory
may undershoot relative to the data is to be expected. Indeed, the CB, in
its rest system, is subject to a flux of high energy nuclei and electrons.
While the electrons are being thermalized, they should contribute a
nonthermal high-energy tail of photons emitted via the ``free-free''
process. Such a power-law tail in an otherwise
approximately-thermal emission is observed from
young supernova remnants (see, e.g., Dyer et al. 2000) and clusters of
galaxies (e.g., Fusco-Femiano et al. 1999; Rephaeli et al., 1999; 
Fusco-Femiano et al., 2000), both of which are systems
 wherein a dilute plasma at a temperature of
$\cal{O}$(1 keV) is exposed to a flux of high energy cosmic rays.

\subsection{Time dependence}

In Fig.(\ref{425}) we compare the single-pulse
light curve of GRB 980425 (Kippen 2000)  with the CB theory, obtained by
integrating Eq.(\ref{boostthermal}) over energy, in the
50-300 keV domain. The parameters used
are $\rm\overline\beta_{in}=1/3$, $\rm\overline\beta_{out}=1/2$,
$\rm\overline{M}_S=1$, $\rm\overline{R}_S=2$, 
$\rm\overline{E}_{CB}=10$ (corresponding to our reference
jet energy in a single pulse), $\rm\overline{\gamma}_{out}=1/3$,
$\rm\overline{\theta}_S=60$, $\rm n=8$ and $\rm z=0.0085$.
For these parameters the GRB fluence, as estimated via 
Eq.(\ref{dfdomega}), is the observed $0.44\times 10^{-5}$ erg cm$^{-2}$. 
Notice that the value used for the viewing angle $\theta$ is very large: 
this is the explanation  (Dar and De R\'ujula 2000a)
why this particular GRB has a normal fluence, in spite of how close its 
progenitor (SN 1998bw) is to us. 

The comparison made in Fig.(\ref{425}) is entirely satisfactory. The parameter
domain giving rise to a light curve with a particular shape, height and width
is much more restricted than the corresponding domain for an energy 
distribution. Yet, we cannot entirely trust the approximate parameter
values thus extracted, for the reasons to be discussed in the next two
subsections.

\subsection{The time-energy correlation}

In Fig.(\ref{712}) we compare the single-pulse
light curves of GRB 990712 (Mallozzi 2000) with the CB theory
(the continuous red curves), obtained by
integrating Eq.(\ref{boostthermal}) over energy, in the
same domains as the data: 20-50 keV (BATSE channel 1.1), 50-100 keV (2.2),
100-300 keV (3.3) and $> 300$ keV (4.4). The parameters used
are $\rm\overline\beta_{in}=1$, $\rm\overline\beta_{out}=1/3$,
$\rm\overline{M}_S=1/4$, $\rm\overline{R}_S=3$, 
$\rm\overline{E}_{CB}=50$ (corresponding to five times our reference
jet energy in a single pulse), $\rm\overline{\gamma}_{out}=1/5$,
$\rm\overline{\theta}_S=1/2$, $\rm n=3$, and $\rm z=0.4315$.

The comparison made in Fig.(\ref{712}) is rather unsatisfactory, in that
the correlation between energy-interval and pulse-width is weaker
in the observations than it is in the predictions. The theoretical
correlation, for a thermal input spectrum, is roughly that implied
by the simplified expression Eq.(\ref{simple}), that is
$\rm \Delta t\;\Delta E\sim H$, in an obvious notation. 
The dashed blue curves in Fig.(\ref{712}) correspond to a modified input
in which we have assumed that the CB cooling
(as discussed in Section 8.5) may be not be linear in time,
but closer to quadratic, so that 
$\rm \Delta t\propto 1/\sqrt{\Delta E}$. This modification
goes in the right direction, but it is still not entirely
satisfactory. We
have not yet investigated in detail how a deviation from
an input thermal spectrum at high energies --that we discussed
in Section 8.1 in commenting Fig.(\ref{123})-- affects the
time-energy correlation. But, since a non-thermal
high-energy tail broadens the energy-distribution
at all times, it ought to weaken 
even further the time-energy correlation, as required.

We have studied the time-energy correlation for other single-pulse GRBs,
such as 981022, 981221 and 990102. They all have a weaker 
energy-interval to pulse-width correlation than our model
predicts, though the problem is most acute for GRB 990712,
that we have thus chosen to expose the limitations of the CB model
in its current formulation.

\subsection{Lessons from the comparison with data}

We conclude from our study of the general properties of GRBs in Section 7,
and from the three comparisons with data in Section 8, that we may
have deliniated the correct overall energetics of the collision
of the CB with the SN shell, but our treatment of the time evolution
of the processes of heating and cooling is oversimplified.
A posteriori, there are many obvious reasons why this ought
to be the case: the front of the CB is no doubt at a higher temperature
than its bulk, since the CB is many collision-lengths long and
is dominantly heated at the front: only muons and their decay
electrons --but not photons from $\pi^0$ decay-- heat the bulk.
The process is not a sudden heating followed by continuous
cooling, as we assumed. We have included cooling by expansion,
but not by emission from the CB's surface. We have assumed
a constant expansion velocity, and not attempted to compute
an actual expansion history from plasma dynamics. The CB
may not have a constant density, it may even be a discontinuous
ball of ``shrapnel''. Etc. etc.

\subsection{An alternative simplified model}

The six general properties of GRBs discussed in Section 7
ought to be quite independent of the complex details
of the CB's collision with the SN shell, since they only
capitalize on the overall energetics and on the fact that,
as it reaches the transparent outskirts of the SN shell,
the CB is cooling by radiation and expansion.
We illustrate this point by sketching an alternative model of
CB heating and cooling, a simplified ``surface'' model that is in
some sense the extreme opposite to the simplified ``volume''
model we have discussed in  detail.
To lighten the discussion, in all numerical results
in this chapter we fix the parameters to their reference values of Table I.

In its rest frame, the front surface of the CB is bombarded by the nuclei
of the SN shell, which have an
energy $\rm m_p \,c^2\,\gamma\sim$ 1 TeV per nucleon, 
roughly 1/3 of which (from $\pi^0\to\gamma\gamma$ decays)
is converted into these
$\gamma$-rays within $\rm X_p\approx m_p/\sigma_{in}(pp)\approx 50\, g\, 
cm^{-2}$, where $\rm \sigma_{in}(pp)$ is the nucleon-nucleon
inelastic cross section.
These high energy photons initiate electromagnetic cascades that  
eventually convert their energy to thermal energy within the CB. 
The radiation length of high energy $\gamma$'s in hydrogenic plasma, 
dominated by $\rm e^+\,e^-$ pair production, is $\rm X_{\gamma e}
\simeq 63$ g cm$^{-2}$,
comparable to $\rm X_p$. The radiation length of thermalized
photons in a hydrogenic plasma is
$\rm X_\gamma^{ion}\approx m_p/\sigma_{_T}\approx 2.6$ g cm$^{-2}$. 

Assume that the quasi-thermal emission rate from
the CB, within $\rm X_\gamma^{ion}$ from its
surface, is in dynamical equilibrium with the fraction of energy deposited
by the CB's collision with the SN shell in that outer layer.
The temperature of the CB's front is then
roughly given by:
\begin{equation}
\rm T(x)\simeq \left[{(n\! -\! 1)\,X_\gamma\, m_p\, c^3 [\gamma(x)]^2\, 
              \sigma_{in}(pp)  \over 
              6\,\sigma\, x_{tp}\,X_{\gamma e}
\, \sigma_{_T}^2}\right ]^{1\over 4}
              \left[ {x\over x_{tp}}\right]^{-{n\over 4}}\!\!\! , 
\end{equation}
where $\rm\gamma(x)$ is a function that decreases
monotonically from $\rm \gamma_{in}$
to $\rm\gamma_{out}$.
Remarkably, only the Lorentz factor of the CBs, but neither their mass 
nor their energy,  appear in the above expression,
except for the fact that, for the result to be correct, they must be
large enough for the CB to pierce the SN shell and remain relativistic.

For $\rm n=8$ the value of  
$\rm x_{tp}$ is $\rm \approx 3 \,R_S$, and, for t close
to $\rm t_{tp}$ or later:   
\begin{equation}
\rm T(t)\simeq 0.16\, keV\,
        \left[ {t_{tp}\over t}\right]^{2}\,\left[ {\gamma(t)\over 
10^3}\right]^{1\over 2}\; . 
\label{newtemp}
\end{equation}
At the time of transparency this is quite comparable to the result of combining 
Eqs.(\ref{Ttrans1}) and (\ref{timedeps}). However, only for $\rm n=4$
does the temperature decrease approximately
as 1/t. For n $>4$ it diminishes faster than 1/t and for $\rm n=8$
it decreases faster than
$\rm 1/t^2$, the ``faster'' being due, in both cases, to the effect
of a decreasing $\rm \gamma(t)$. For an exact $\rm 1/t^2$ behaviour 
the pulse width narrows with time as $\rm \Delta t \propto E^{-0.5}$ and,
as we have also seen in
Section 8.3, this goes in the direction of improving the
predicted time-energy correlation. In fact, Fenimore et al.~(1995)
found, from a large sample of GRB pulses, that 
$\rm \Delta t \propto E^{-0.46}$.

The total radiated energy, in the CB rest frame, is roughly the thermal
energy deposition within one radiation length from its
front surface. After attenuation in the SN shell, it reduces to:
\begin{equation}
\rm  E_{pulse}^{rest}\approx  {\sigma_{in}(pp)\, \pi\, 
[R_{CB}^{tp}]^2\,\bar X_\gamma\, m_p\,c^2\, \gamma(t)
                      \over 3\,X_{\gamma e}\, \sigma_{_T}^2}\,,
\label{newenergy}  
\end{equation}
where  $\rm  \bar X_\gamma$ is the radiation length  
in the obscuring shell averaged over the black body spectrum. 
For a typical $\gamma$-ray
peak energy of $\rm E_p\sim 1\, MeV$ in the SN rest frame,
$\rm \bar X_\gamma\simeq 10\ g\, cm^{-2}\,. $ Consequently,
the CB's radius at transparency is
 $\rm R_{CB}^{tp}=4\times 10^{11}$ cm and
$\rm  E_{pulse}^{rest}\sim 3\times 10^{45}\, erg$,
for $\rm\gamma(t)\sim 10^3$.  This is consistent with the results in
Eqs.(\ref{EGRB1}), implying that the predicted fluences in the
surface-heating and volume-heating models are quite similar.
The fact that the characteristic
temperatures of the volume-heating and the surface-heating
models around the time of transparency
are also similar means that their predicted GRB 
individual-photon energies are comparable and both in agreement
with the GRB observations.

\section{Conclusions}

In a previous paper (Dar and De R\'ujula 2000a) we have argued
that the CB model provides a very good description of GRB
afterglows, including those whose light curve is seen to
rise before it drops, as is the case for GRB 970508. There
we also contended that, in the case of GRB 980425, the model
provides a strong motivation for the search of the superluminal
motion of the afterglow-emitting CB, relative to the associated
supernova: SN1998bw. This would be a decisive signature
for highly relativistic cannonballs, as opposed to conically
spreading jets. We plan to discuss in future work other important signatures
of the CB model: high energy neutrinos and photons during the GRB, 
flare up and X-ray lines in its early afterglow.

In this paper we have demonstrated that the CB model 
explains the fluence and energy spectrum of GRBs,
as well as the characteristic properties of their light curves.
The detailed heating, expanding and
cooling of the CB ---as it hits and sweeps up the SN shell---
we have treated only in a simplified fashion. As a consequence,
the model in its present form does not provide a completely
satisfactory quantitative description of the time-energy correlation.
We may not have completely untied the perduring Gordian
knot of the GRB conundrum, but we have argued that
we have sliced it open.
\vspace{.5cm}

\noindent
{\bf Aknowledgements} 
We are indebted to Rainer Plaga for interesting discussions.
This work was supported in part by the Fund for Promotion Of Research At 
The Technion.

\vskip 1 true cm
%\vskip 0.2 true cm
\begin{table}[h]
%\vskip 0.1 true cm
%\hspace{-.5cm} %if you want to center your table act on this argument
\hspace{0.4 cm}
\begin{tabular}{|l|c|c|c|}
\hline
\hline
$\;\;\;\;\;\;\;\;\;$Parameter   &Symbol &Value \\
\hline
SN-shell's mass     & $\rm M_S$             & $\rm 10\; M_\odot$ \\
SN-shell's radius   & $\rm R_S$              & $2.6\times 10^{14}$ cm \\
SN-shell's density index   &n  &8 \\
Outgoing Lorentz factor   & $\rm\gamma_{out}$  & $10^3$ \\
CB's viewing angle  & $\theta$ & $\rm 10^{-3}$  \\
CB's energy   & $\rm E_{CB}$ & $10^{52}$ erg   \\
Initial $\rm v_{_T}/c$ of expansion & $\rm\beta_{in}$ & $1/(10\,\sqrt{3})$ \\
Final $\rm v_{_T}/c$ of expansion & $\rm\beta_{out}$ & $1/\sqrt{3}$ \\
\hline
Redshift   &z  &1   \\
\hline
\hline
\end{tabular}
\end{table}
\vskip -0.3 true cm
\noindent
{\bf Table I.}
List of the ``reference'' values of various parameters. In the text a barred
parameter means its actual value divided by its reference value, so that,
for instance, $\rm \overline M_S=1/2$ means the actual mass of the SN shell
is taken to be $\rm 5\; M_\odot$.

{}

%%%%%%%%%%%%%%%%%%%%%%%%%%%%1
\begin{figure}
\begin{center}
\vspace*{1.0cm}
\hspace*{-.5cm}
\epsfig{file=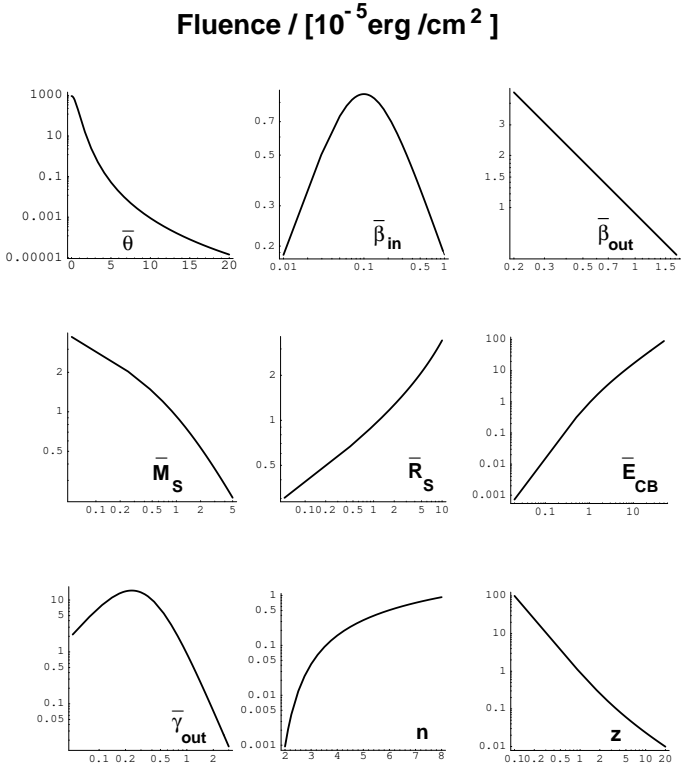,width=9cm}
%\vspace*{-14.6cm}
\caption{Fluence as a function of  various parameters,
in units of $10^{-5}$ erg/cm$^2$. The parameters
$\theta$ through $\rm\gamma_{out}$ are in units
of their reference values of Table I, thus the barred notation.
The index n and the redshift z are not rescaled.
All parameters not being varied are fixed at their
reference values of Table I, but for $\theta$, fixed
at a ``typical'' $\rm \overline{\theta}=3.$}
\vspace*{-0.5cm}
\label{flu}
\end{center}  
\end{figure}
%%%%%%%%%%%%%%%%%%%%%%%%%%%%2
\begin{figure}
\begin{center}
\vspace*{1.0cm}
\hspace*{-1cm}
\epsfig{file=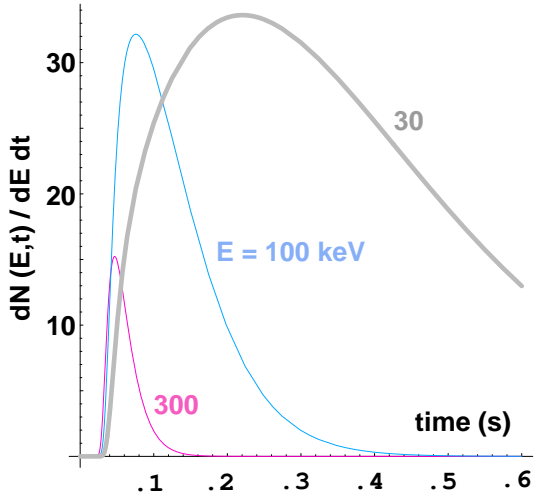,width=7cm}
%\vspace*{-14.6cm}
\caption{GRB-pulse shape as a function of time, at various
fixed $\gamma$-energies.}
\vspace*{-0.5cm}
\label{t3Es}
\end{center}  
\end{figure}
%%%%%%%%%%%%%%%%%%%%%%%%%%%%

%%%%%%%%%%%%%%%%%%%%%%%%%%%%3
\begin{figure}
\begin{center}
\vspace*{1.0cm}
\hspace*{-1cm}
\epsfig{file=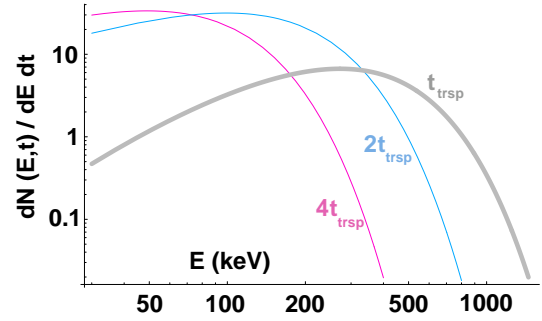,width=7cm}
%\vspace*{-14.6cm}
\caption{GRB $\gamma$-energy distributions,
at various fixed times, multiples of the (observer's)
time at which the SN-shell becomes transparent.}
\vspace*{-0.5cm}
\label{E3ts}
\end{center}  
\end{figure}
%%%%%%%%%%%%%%%%%%%%%%%%%%%%

%%%%%%%%%%%%%%%%%%%%%%%%%%%%4
\begin{figure}
\begin{center}
\vspace*{1.0cm}
\hspace*{-1cm}
\epsfig{file=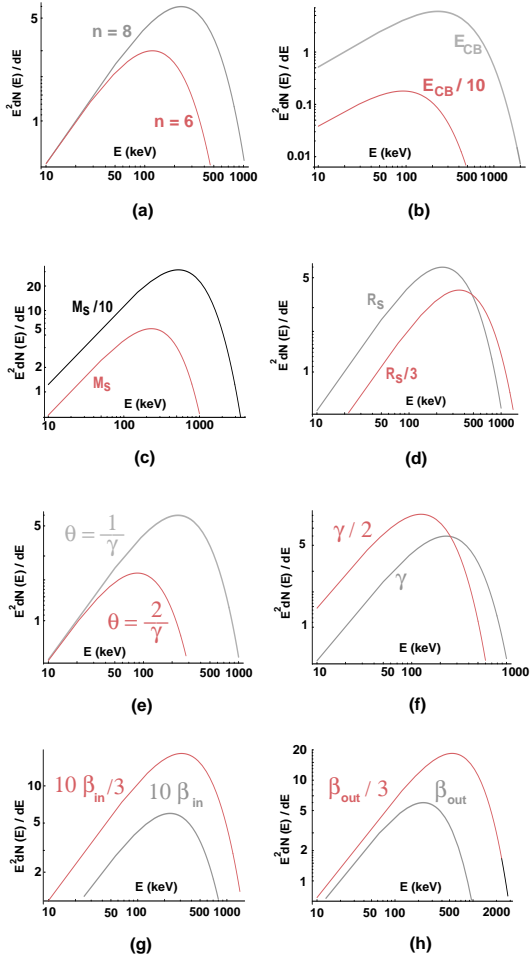,width=7cm}
%\vspace*{-14.6cm}
\caption{Time-integrated $\rm E^2\,dN/dE$ distributions,
illustrating the sensitivity to one parameter at a time.
The absolute vertical scale is arbitrary, but the relative scales
are not. All parameters not mentioned in each subfigure are
kept at the reference values of Table I
(but for $\rm\beta_{in}$, which is fixed at $1/\sqrt{3}$).
 Notice that the shape of the energy
distribution is always the same, irrespective of the
parameter values.}
\vspace*{-0.5cm}
\label{E}
\end{center}  
\end{figure}
%%%%%%%%%%%%%%%%%%%%%%%%%%%%
%%%%%%%%%%%%%%%%%%%%%%%%%%%%5
\newpage
\begin{figure}
\begin{center}
\vspace*{1.0cm}
\hspace*{-.5cm}
\epsfig{file=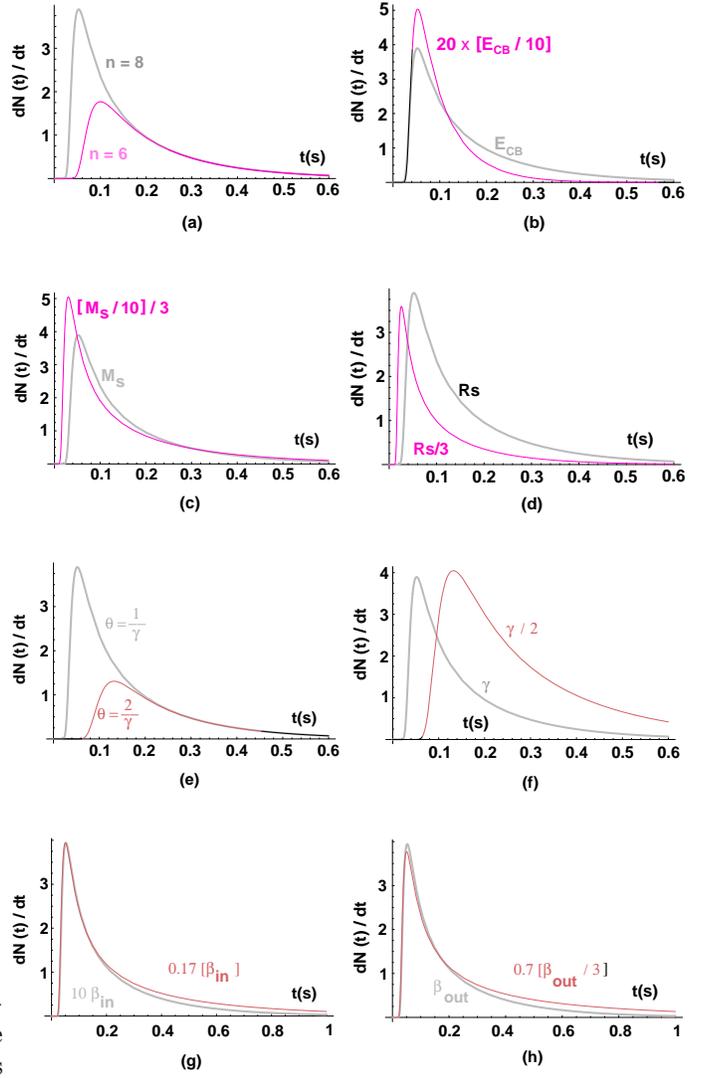,width=9cm}
%\vspace*{-14.6cm}
\caption{Time distribution $\rm dN/dt$ of a single GRB pulse,
integrated for all $\rm E>30$ keV. The absolute vertical scale is 
arbitrary, but the relative scales are not. All parameters not mentioned 
in each subfigure are kept at the reference values of Table I
(but for $\rm\beta_{in}$, which is fixed at $1/\sqrt{3}$).
For ease of comparison, in some subfigures, a curve has been rescaled,
e.g. in (b) the result for an input $\rm E_{CB}/10$ has been multiplied by 20.}
\vspace*{-0.5cm}
\label{t}
\end{center}  
\end{figure}
%%%%%%%%%%%%%%%%%%%%%%%%%%%%

%%%%%%%%%%%%%%%%%%%%%%%%%%%%6
\newpage
\begin{figure}
\begin{center}
\vspace*{1.0cm}
\hspace*{-1cm}
\epsfig{file=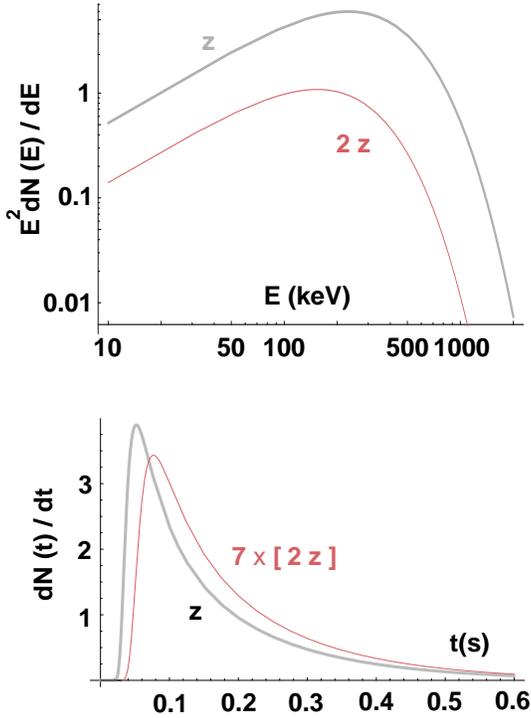,width=7cm}
%\vspace*{-14.6cm}
\caption{Illustration of the sensitivity to redshift, z, of the
energy-and time-distributions in a GRB pulse.}
\vspace*{-0.5cm}
\label{z}
\end{center}  
\end{figure}
%%%%%%%%%%%%%%%%%%%%%%%%%%%%
%%%%%%%%%%%%%%%%%%%%%%%%%%%%7
\begin{figure}
\begin{center}
\vspace*{1.0cm}
\hspace*{-1cm}
\epsfig{file=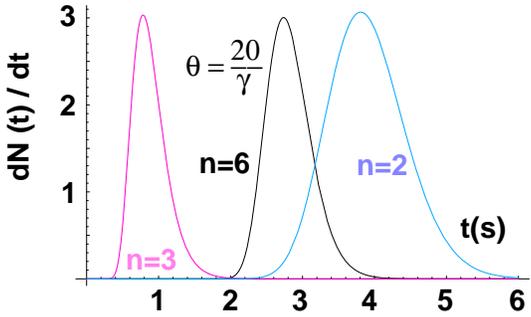,width=7cm}
%\vspace*{-14.6cm}
\caption{Examples of parameter values that give rise to
non-FRED pulse shapes. The individual vertical
scales are chosen for ease on comparison. 
All parameters not mentioned are
as in Table I.}
\vspace*{-0.5cm}
\label{t2}
\end{center}  
\end{figure}
%%%%%%%%%%%%%%%%%%%%%%%%%%%%
%%%%%%%%%%%%%%%%%%%%%%%%%%%%8
\newpage
\begin{figure}
\begin{center}
\vspace*{1.0cm}
\hspace*{-1cm}
\epsfig{file=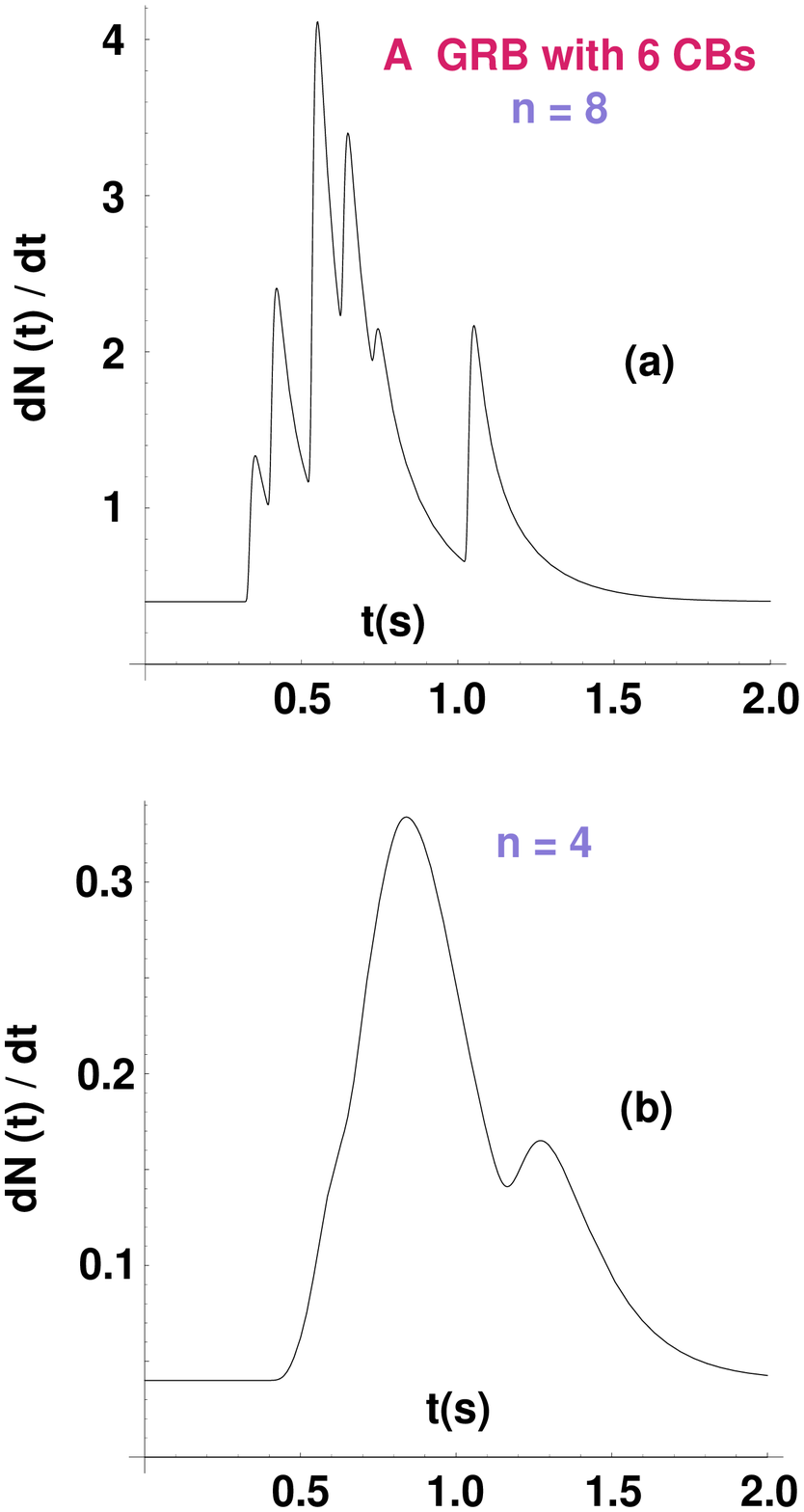,width=9cm}
%\vspace*{-14.6cm}
\caption{ ``Synthetic'' GRB light curves, generated by shooting
six CBs at random in a 1.5 s time-interval, and with random values 
of $\rm E_{CB}$ within a factor of three of our reference value.
The only difference between (a) and (b) is that $\rm n = 8$ in (a),
while $\rm n = 4$ in (b). All other parameters in this figure have their 
reference values. The figure
illustrates how a CB produces a GRB pulse, but a
GRB-pulse may not correspond to a single CB.}
\vspace*{-0.5cm}
\label{6CB}
\end{center}  
\end{figure}
%%%%%%%%%%%%%%%%%%%%%%%%%%%%

%%%%%%%%%%%%%%%%%%%%%%%%%%%%9
\begin{figure}
\begin{center}
\vspace*{1.5cm}
\hspace*{-1cm}
\epsfig{file=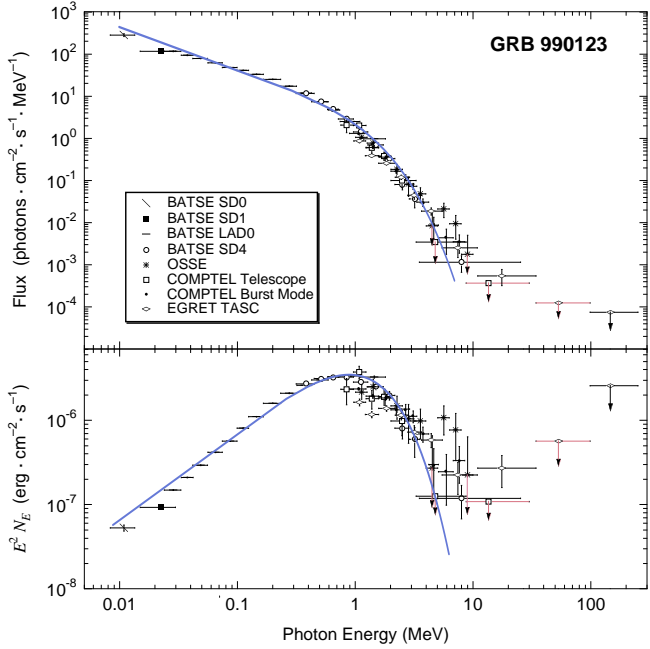,width=9cm}
%\vspace*{-14.6cm}
\caption{Comparison of theory and observation
for the time-integrated energy distributions $\rm dN/dE$
and $\rm E^2\,dN/dE$, in the case of GRB 990123.
Notice that many experimental points at the
higher energies are only upper limits.}
\vspace*{-0.5cm}
\label{123}
\end{center}  
\end{figure}
%%%%%%%%%%%%%%%%%%%%%%%%%%%%
%%%%%%%%%%%%%%%%%%%%%%%%%%%%10
\begin{figure}
\begin{center}
\vspace*{1.0cm}
\hspace*{-1cm}
\epsfig{file=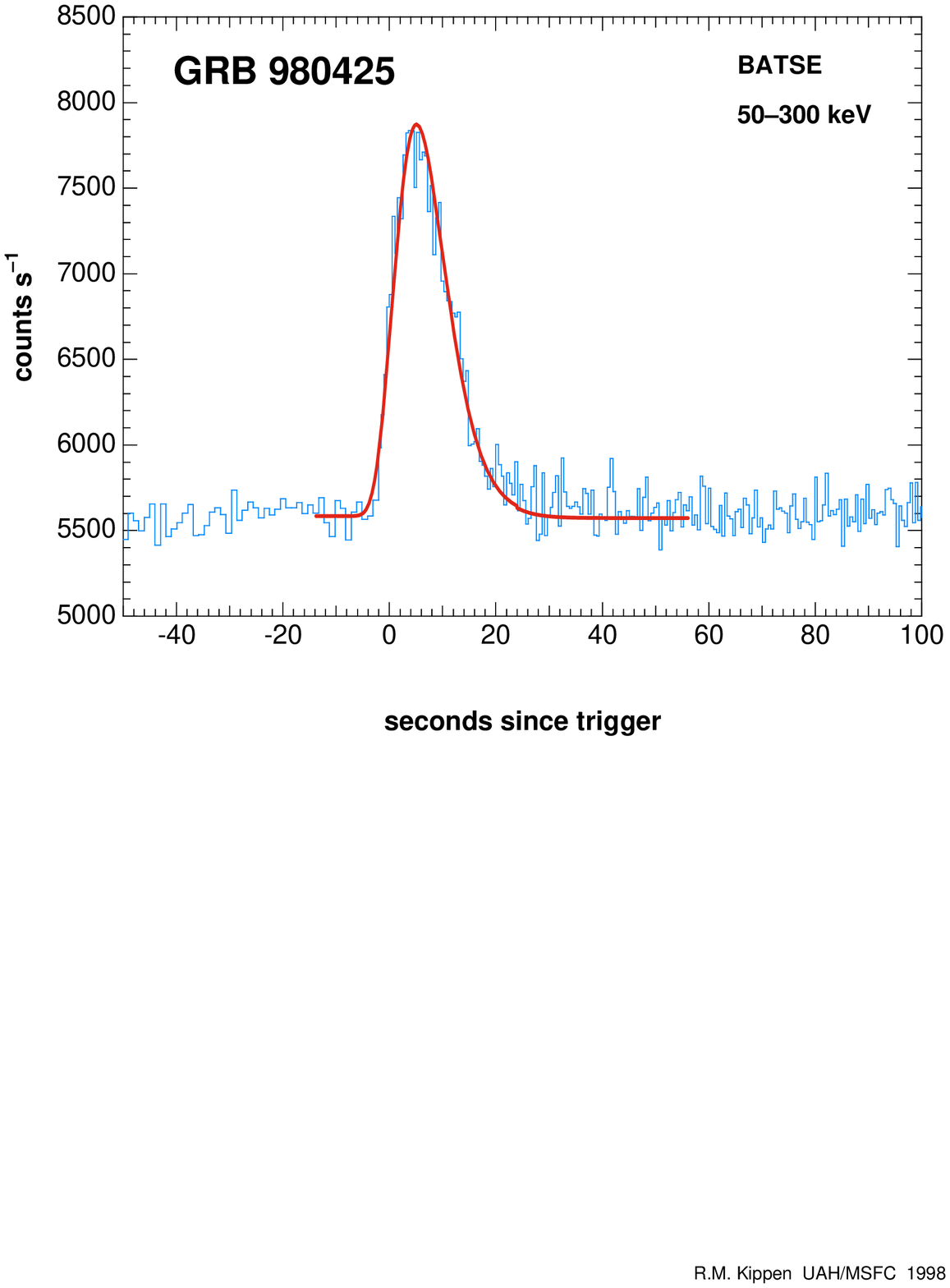,width=8cm}
%\vspace*{-14.6cm}
\caption{Comparison of theory and observation for the
light curve of GRB 980425, in the 50-300 keV energy interval. }
\vspace*{-0.5cm}
\label{425}
\end{center}  
\end{figure}
%%%%%%%%%%%%%%%%%%%%%%%%%%%%
%%%%%%%%%%%%%%%%%%%%%%%%%%%%11
\begin{figure}
\begin{center}
\vspace*{1.0cm}
\hspace*{-.5cm}
\epsfig{file=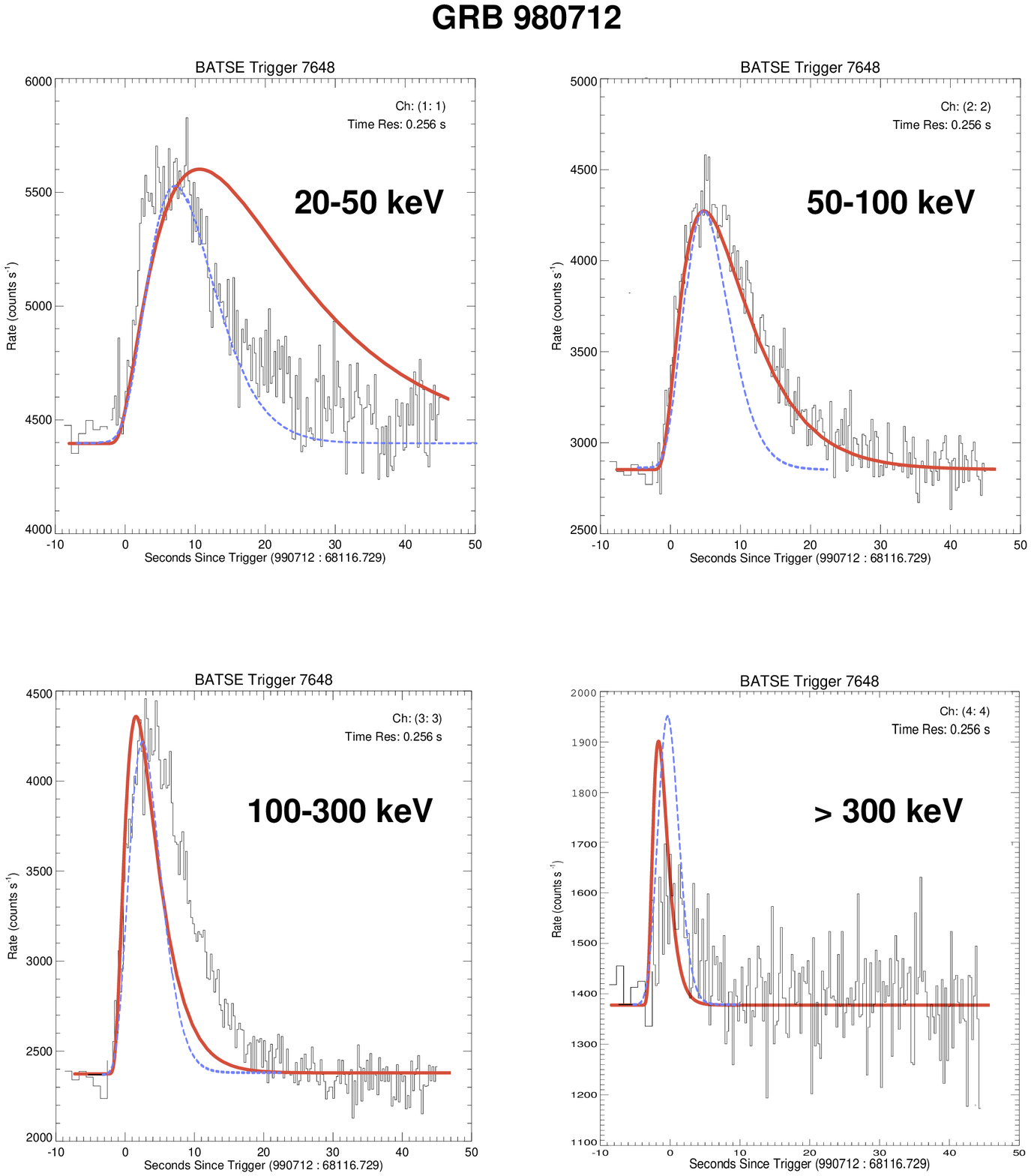,width=9cm}
%\vspace*{-14.6cm}
\caption{Comparison of theory and observations for the
light curve of GRB 980712, in various BATSE energy intervals. }
\vspace*{-0.5cm}
\label{712}
\end{center}  
\end{figure}
%%%%%%%%%%%%%%%%%%%%%%%%%%%%

\end{document}